\makeatletter
\let\kernel@label\label
\makeatother

\documentclass[aps,prb,12pt,floatfix,longbibliography,a4paper]{revtex4-2}

\makeatletter
\let\label\kernel@label
\makeatother

\usepackage{graphicx}
\usepackage{amssymb}
\usepackage{longtable}
\usepackage{amsmath,bm}
\usepackage{xcolor}
\usepackage{physics}
\usepackage[normalem]{ulem}
\usepackage{array}
\usepackage{makecell}

\usepackage{placeins}

\makeatletter
\@addtoreset{figure}{hoge}
\makeatother
\makeatletter
\@addtoreset{equation}{hoge}
\makeatother
\makeatletter
\@addtoreset{section}{hoge}
\makeatother
\makeatletter
\@addtoreset{table}{hoge}
\makeatother
\makeatletter
\@addtoreset{equation}{hoge}
\makeatother

\usepackage{hyperref}

\newcommand{\be}{\begin{equation}}
\newcommand{\ee}{\end{equation}}
\newcommand{\bfig}{\begin{figure}}
\newcommand{\efig}{\end{figure}}

\newcommand{\RRS}{\textit{R}Ru$_3$Si$_2$}

\newcommand{\LRS}{LaRu$_3$Si$_2$}
\newcommand{\CRS}{CeRu$_3$Si$_2$}

\newcommand{\NRS}{NdRu$_3$Si$_2$}

\begin{document}
\title{Bond-density-wave orders induced by \\ geometric frustration in the kagome metal \CRS{}}

\author{Ryo Misawa$^{1}$}\email{misawann6@g.ecc.u-tokyo.ac.jp}
\author{Shunsuke Kitou$^{2}$}
\author{Rinsuke Yamada$^{1}$}
\author{Xiaolong Feng$^{3}$}
\author{Ryota Nakano$^{1}$}
\author{Priya Ranjan Baral$^{1}$}
\author{Yuiga Nakamura$^{4}$}
\author{Leslie M. Schoop$^{5}$}
\author{Yukitoshi Motome$^{1}$}
\author{Taka-hisa Arima$^{2,6}$}
\author{Xiuzhen Yu$^{6}$}
\author{Max Hirschberger$^{1,6}$}\email{hirschberger@ap.t.u-tokyo.ac.jp}

\affiliation{$^{1}$Department of Applied Physics and Quantum-Phase Electronics Center (QPEC), The University of Tokyo, Bunkyo-ku, Tokyo 113-8656, Japan}
\affiliation{$^{2}$Department of Advanced Materials Science, The University of Tokyo, Kashiwa, Chiba 277-8561, Japan}
\affiliation{$^{3}$Max Planck Institute for Chemical Physics of Solids, 01187 Dresden, Germany}
\affiliation{$^{4}$Japan Synchrotron Radiation Research Institute (JASRI), SPring-8, Hyogo 679-5198, Japan}
\affiliation{$^{5}$Department of Chemistry, Princeton University, Princeton, New Jersey 08544, USA}
\affiliation{$^{6}$RIKEN Center for Emergent Matter Science (CEMS), Wako, Saitama 351-0198, Japan}

\maketitle
\newpage

\begin{center}
\Large{Abstract}
\end{center}
\textbf{Geometric frustration gives rise to vast manifolds of degenerate ground states and competing orders in spin and charge systems. Typically, classical ground states are governed by a local ``zero-sum constraint" that relieves frustrated antiferromagnetic interactions or Coulomb repulsion. To date, the paradigm of geometric frustration has yielded a rich landscape of emergent phases, from spin ices and quantum spin liquids to charge glasses. However, an analogous phase rooted in chemical bonding has yet to be firmly demonstrated. Here we report the discovery of bond-density-wave orders induced by geometric frustration in the kagome metal \CRS{} above room temperature. Through synchrotron X-ray diffraction, real-space transmission electron microscopy, and model calculations, we observe two distinct long-period superlattices with harmonic and anharmonic structural modulations. Crucially, interlayer bonds between kagome planes modulate in a sublattice-selective manner to fulfill the zero-sum constraint on the kagome lattice. We demonstrate the potential of kagome metals to host complex bond-ordered states constrained by geometric frustration and establish chemical bonding as a distinct pathway to frustration physics in quantum materials even above room temperature.
}

\newpage

\begin{center}
\Large{Main Text}
\end{center}

\begin{figure*}[ht]
  \begin{center}
	\includegraphics[clip, trim=0cm 0cm 0cm 0cm, width=0.8\linewidth]{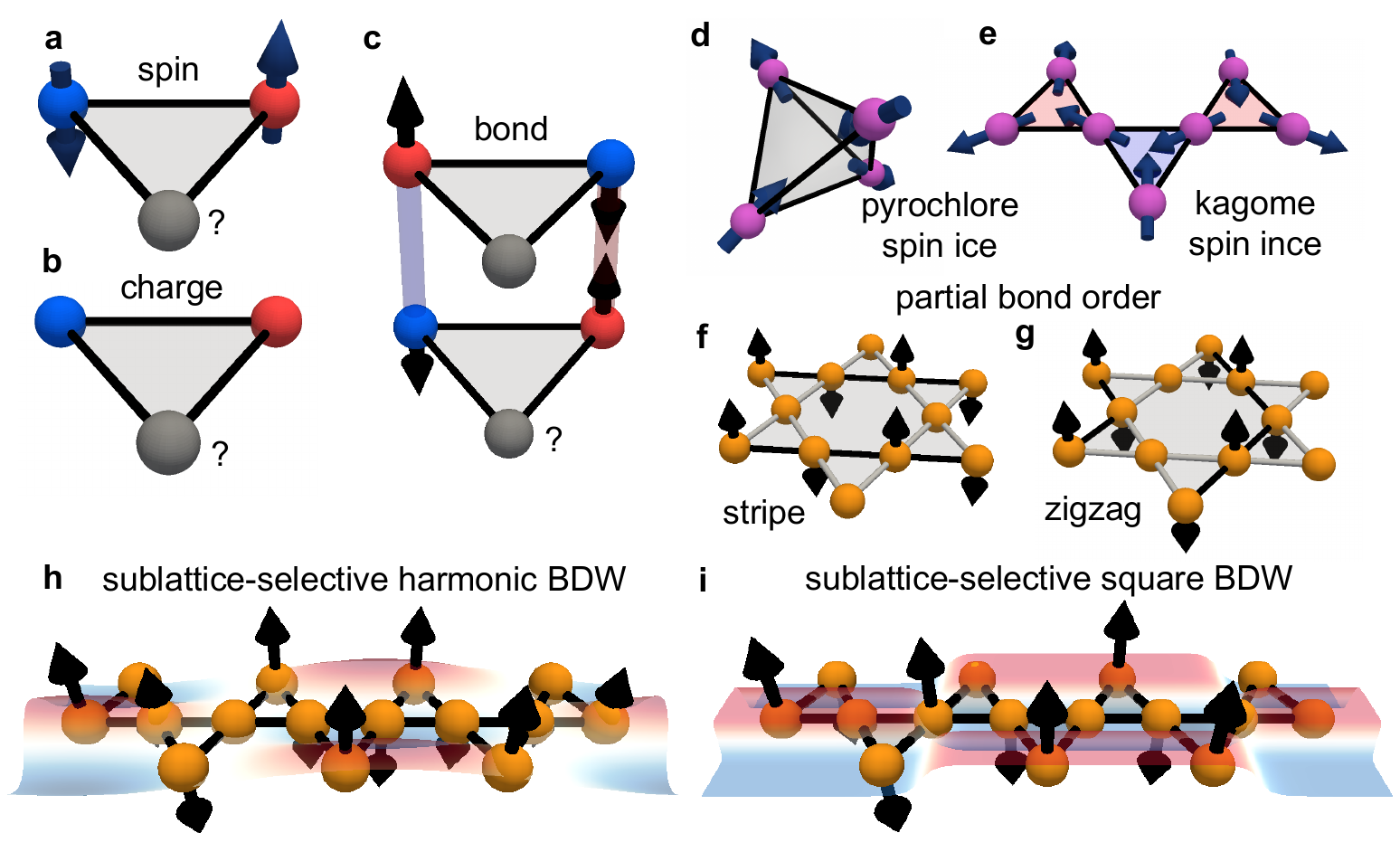}
    \caption[]{\textbf{Bond-ordered states satisfying a zero-sum constraint on a geometrically frustrated lattice.} \textbf{a},\textbf{b},\textbf{c}, Frustration of spin, charge, and interlayer bond on the triangular lattice with antiferromagnetic interactions or Coulomb repulsion. The red (blue) sphere represents up (down) spin, positive (negative) charge, or a site bonded with the upper (lower) layer. The navy (black) arrow marks the spins (interlayer bond direction), while the colored cylinders represent chemical bonding.
    \textbf{d}, Representative two-in–two-out spin configuration on the pyrochlore lattice, a local motif of spin ice that satisfies a zero-sum constraint. \textbf{e}, Kagome spin ice with one-in-two-out (positive ``charge", red) and two-in-one-out (negative ``charge", blue) clusters. \textbf{f},\textbf{g}, Partial bond order on the kagome lattice with stripe and zigzag patterns of interlayer bonding. Two of the three sublattices form bonds with opposite layers, while the other does not, to realize neutral bonding on each triangle. \textbf{h},\textbf{i}, Sublattice-selective harmonic and anharmonic square bond-density-wave (BDW) orders on the kagome lattice. As in partially bond-ordered states, the local bond zero-sum constraint is satisfied by two out-of-phase waves (colored mesh), minimizing frustrated Coulomb repulsion.
     }
    \label{fig:fig1}
  \end{center}
\end{figure*}

\textbf{Introduction}\\
Frustration is defined by the inability to simultaneously minimize all local interaction energies, often leading to a highly degenerate manifold of low-energy collective states. In crystalline solids, frustration is most commonly enforced by lattice geometry~\cite{Ramirez1994-ac}. A canonical example is antiferromagnetically coupled Ising spins on a triangular lattice (Fig.~\ref{fig:fig1}\textbf{a}), where geometric constraints prevent the simultaneous minimization of all pairwise interactions. Closely related physics appears in mixed-valence systems, in which inhomogeneous charge distributions experience frustrated Coulomb repulsion between charge-rich and charge-poor sites (Fig.~\ref{fig:fig1}\textbf{b}). Across these settings, frustration has underpinned exotic phases from quantum spin liquids~\cite{Anderson-QSL,Pauling-RVB,Kitaev2006-em,Takagi2019-mq} and topological spin textures~\cite{Okubo2012-ch} to charge-glass states~\cite{Kagawa2013-zt,Sasaki2017-la}.

Typically, classical ground states on the triangle lattice are determined by the ``zero-sum constraint", which encodes an energetic preference for local neutrality. For Heisenberg spins $\bm{S}_i$ interacting via nearest-neighbour antiferromagnetic exchange, the Hamiltonian is
\begin{equation}
\label{eq:1}
\mathcal{H} = J\sum_{\langle i, j\rangle}\bm{S}_i\cdot\bm{S}_j
= \frac{J}{2}\sum_{\bigtriangleup}(\bm{S}_1 + \bm{S}_2 + \bm{S}_3)^2 + \mathrm{const.},
\end{equation}
where $\triangle$ denotes a summation over upward triangles. Ground states therefore minimize $(\bm{S}_1 + \bm{S}_2 + \bm{S}_3)^2$, or suppress an effective cluster ``charge''. This neutrality constraint on elementary clusters, beyond triangular motifs, parallels Pauling’s ice rule~\cite{Pauling-ice} and the Anderson condition~\cite{Anderson-CO}, and underlies the emergence of spin-ice in pyrochlore (Fig.~\ref{fig:fig1}\textbf{d})~\cite{Ramirez1999-ct,Pomaranski2013-ao} and in kagome magnets (Fig.~\ref{fig:fig1}\textbf{e})~\cite{Zhao2020-da}, as well as charge order in spinel compounds~\cite{Senn2011-pb,Horibe2006-mf,Okamoto2020-ju}.

Here we extend the concept of geometric frustration to chemical bonding, in which the formation of anisotropic interlayer bonds can mimic Ising antiferromagnets (Fig.~\ref{fig:fig1}\textbf{c}). In this picture, the charge $(\bm{S}_1 + \bm{S}_2 + \bm{S}_3)^2$ is interpreted as the cluster-averaged interlayer bond length, and the zero-sum constraint is satisfied by bonding patterns that remain neutral on each cluster (bond zero-sum constraint). The simplest realizations on the kagome lattice are stripe and zigzag arrangements of partially ordered bonds (Fig.~\ref{fig:fig1}\textbf{f},\textbf{g}), in which two of the sublattices form antiferroic bonds, while the other sublattice remains unbonded. This represents a form of partial order analogous to frustrated Ising magnets~\cite{Debauche1991-ma,Mekata1977-uf} and Kondo singlets~\cite{Mentink1994-di,Donni1996-pr,Lacroix1996-yp,Ballou1998-lw}.

Pushing beyond these simplest motifs, we uncover a new class of harmonic and anharmonic bond-density-wave (BDW) orders in a kagome metal, characterized by a modulation of interlayer bonds between kagome planes  (Fig.~\ref{fig:fig1}\textbf{h},\textbf{i}). Viewed structurally, the BDW orders can be classified within a broad class of buckled structures: out-of-plane modulations arising from lattice or bonding instabilities (see Supplementary Section). In contrast to other buckled superstructures~\cite{Plokhikh2024-eg,Mielke2025-nb,Devarakonda2024-fo}, we here refine the long-period superstructures in all phases and identify novel sublattice-selective textures of the bond modulation, which obey the local bond zero-sum constraint emerging from geometric frustration on the kagome lattice, rather than trivially from crystal symmetry.

\begin{figure}[t]
  \begin{center}
		\includegraphics[clip, trim=0cm 0cm 0cm 0cm, width=1.\linewidth]{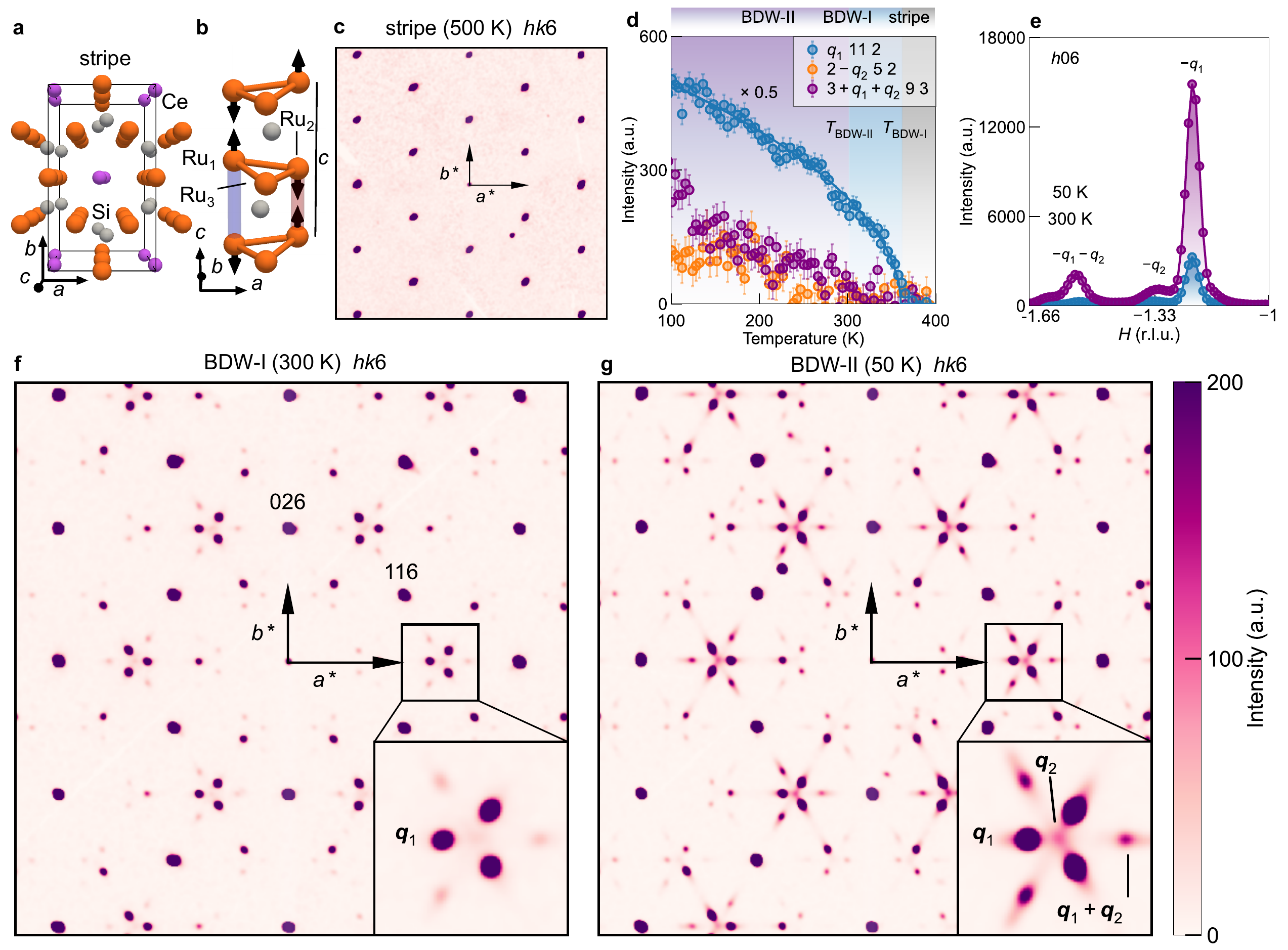}
    \caption[]{\textbf{Observation of harmonic and anharmonic BDW orders in the kagome metal \CRS{}.}
    \textbf{a},\textbf{b}, Top and side views of the parent orthorhombic structure of \CRS{} (stripe, space-group: $Ibmm$). Ru$_1$ and Ru$_2$ form a $d{_z{^2}}$ molecular orbital with opposite kagome layers, whereas Ru$_3$ remains unbonded, realizing partial bond order.
    \textbf{c}, Reciprocal-space maps in the $hk6$ plane reconstructed from synchrotron X-ray diffraction (XRD) data at $500\,$K (stripe).
    \textbf{d}, Temperature dependence of superlattice reflection intensities corresponding to $\bm{q}_1 = (2/9, 0, 0)$ (cyan), $\bm{q}_2 = (1/3, 0, 0)$ (orange), and $\bm{q}_1+\bm{q}_2$ (purple). The intensities for $\bm{q}_1$ are scaled by a factor of $0.5$. Superlattice reflections of BDW-I and BDW-II emerge from $T_\text{BDW-I} \sim 360\,$K and $T_\text{BDW-II} \sim 300\,$K, respectively. A cyan line is a mean-field fit to the data and is proportional to $\sqrt{T-T_\text{BDW-I}}$.
    \textbf{e}, Line cut along the $h06$ line at $300\,$K (cyan) and $50\,$K (purple). The $\bm{q}_1$ superlattice reflection exists at $300\,$K, while its higher-order harmonics $\bm{q}_2 $ and $\bm{q}_1+\bm{q}_2$ are clearly observed at $50\,$K.
    \textbf{f},\textbf{g}, Reciprocal-space maps in the $hk6$ plane at $300\,$K (BDW-I, harmonic), and $50\,$K (BDW-II, anharmonic).
    }
    \label{fig:fig2}
  \end{center}
\end{figure}

\bigskip
\textbf{Partial bond order in kagome metals}\\
As a candidate platform for frustrated bond orders, we study the prototypical kagome metal \CRS{} hosting kagome-derived bands of Ru $4d$ orbital character near the Fermi level~\cite{Rauchschwalbe1984-wk,Kishimoto2003-mp,Mielke2021-ha,Junze2025-pd}. In recent studies~\cite{Plokhikh2024-eg,Misawa2025-adv}, it is shown that, at high temperature, this family of compounds \RRS{} (\textit{R}=rare-earth) forms orthorhombic structures which are characterized by partial ordering of Ru-$d_{z^2}$ molecular orbitals: the interlayer bonds between kagome planes form stripe patterns (Figs.~\ref{fig:fig1}\textbf{f},~\ref{fig:fig2}\textbf{a},\textbf{b}) or zigzag patterns (Fig.~\ref{fig:fig1}\textbf{g}).
In particular, two types of stripe phases have nearly the same energy due to the geometry of the kagome lattice, as evidenced by our phonon calculations (see Supplementary Section), and their competition leads to short-range bond order in \NRS{} as a metastable phase at elevated temperature~\cite{Misawa2025-adv}. As such, this class of materials constitutes an ideal platform for investigating frustrated bond order on the kagome lattice. It is also reported that \LRS{} hosts $\bm{q} = (0, 1/2, 0)$ and $(0, 1/3, 0)$ superstructures, but the ground state is not refined, and the absence of higher-order harmonics precludes the distinction between multi-domain and multi-$\bm{q}$ states~\cite{Plokhikh2024-eg,Mielke2025-nb}.

To first examine the high-temperature structure of \CRS{}, we perform synchrotron X-ray diffraction (XRD) on single crystals. Figure~\ref{fig:fig2}\textbf{c} shows the XRD pattern in the $hk6$ plane of reciprocal space; we use the orthorhombic basis for the Miller indices $hkl$. We identify the parent structure as the stripe phase (Figs.~\ref{fig:fig1}\textbf{f},~\ref{fig:fig2}\textbf{a}); see Supplementary Section for details of structure refinement. Figure~\ref{fig:fig2}\textbf{b} shows the side view of this structure; sublattices Ru$_1$ and Ru$_2$ form interlayer bonds in opposite directions, while Ru$_3$ remains unbonded.

\bigskip
\textbf{Observation of harmonic and anharmonic BDW orders.}\\
Upon cooling below $T_\text{BDW-I}\sim 360\,$K (Fig.~\ref{fig:fig2}\textbf{d}), we observe superlattice reflections in \CRS{} (BDW-I). At $300\,$K, a line profile along the $h06$ line in Fig.~\ref{fig:fig2}\textbf{e} evidences the formation of a superlattice with a commensurate modulation vector, $\bm{q}_1 = (2/9, 0, 0)$. Due to orthorhombic domains related by threefold rotation, the superlattice reflections appear at symmetry-related positions (Fig.~\ref{fig:fig2}\textbf{f}). The superlattice reflections have two essential features: (i) comparable intensities at $q_1$ $k$ $l=2n+1$ with other superlattice reflections, and (ii) presence of superlattice reflections only around forbidden fundamental reflections: $h+k+l=2n+1$ are absent due to body centering of the parent structure (see Supplementary Section). This suggests that long-period structural modulation along the $a$ axis has the character of a transverse optical wave: the atomic displacements are normal to the propagation direction of the wave and are out of phase between neighboring sites (Supplementary Section). 

Further cooling down, we observe additional superlattice reflections in Fig.~\ref{fig:fig2}\textbf{g} (BDW-II) emerging below $T_\text{BDW-II}\sim 300\,$K (Fig.~\ref{fig:fig2}\textbf{d}). Comparing the same line cut at $50\,$K and $300\,$K (Fig.~\ref{fig:fig2}\textbf{e}), we identify $\bm{q}_2 = (1/3, 0, 0) = -3\bm{q}_1 + \bm{G}$ and $\bm{q}_1 + \bm{q}_2 = 7\bm{q}_1-\bm{G}$ with $\bm{G}=(1,0,0)$, which are thus third and seventh order harmonics of $\bm{q}_1$. The emergence of the higher harmonics signals the formation of an anharmonic superlattice.

\begin{figure*}[t]
  \begin{center}
		\includegraphics[clip, trim=0cm 0cm 0cm 0cm, width=0.9\linewidth]{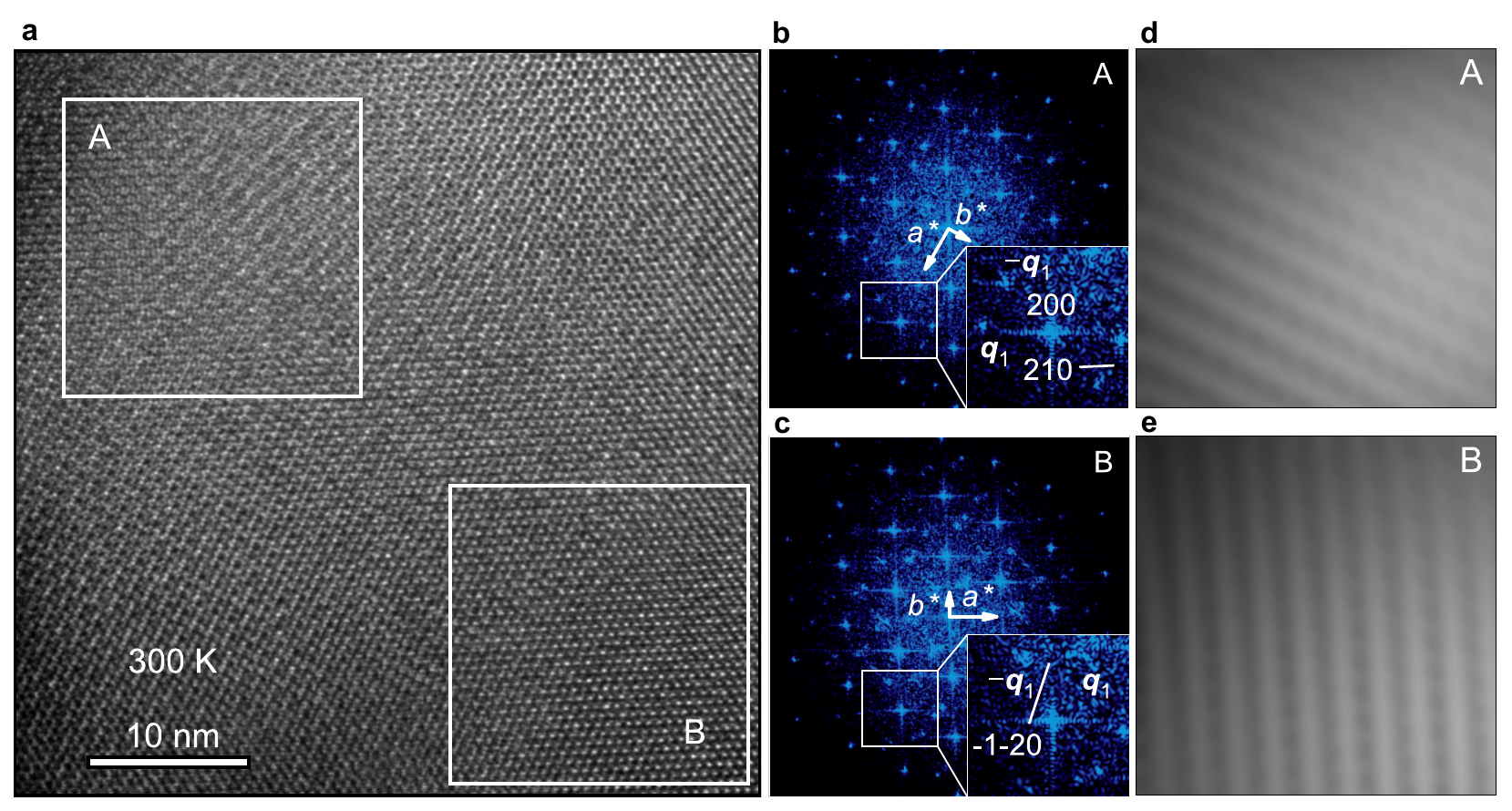}
    \caption[]{\textbf{Real-space observation of the BDW order at room temperature.} \textbf{a}, Real-space image of the (001) surface obtained by transmission electron microscopy (TEM). \textbf{b},\textbf{c}, Fast Fourier transform (FFT) patterns of two domains A and B in panel~\textbf{a}. The arrows indicate reciprocal-space vectors in each orthorhombic domain. Inset: close-up view around a Bragg reflection with superlattice reflections existing only next to extinct Bragg reflections ($2\,1\,0$ and $2\,\mbox{-}1\,0$ for panel~\textbf{b}). \textbf{d},\textbf{e}, Inverse FFT patterns of the domains A and B reconstructed from superlattice reflections (Methods). The amplitude of the real part is displayed after application of a Gaussian filter.
    }
    \label{fig:fig3}
  \end{center}
\end{figure*}

\bigskip
\textbf{Real-space observation of the BDW order at room temperature}\\
To observe the real-space texture of the superlattice, we use transmission electron microscopy (TEM) at room temperature. Figure~\ref{fig:fig3}\textbf{a} shows the high-resolution TEM image. We illustrate in Fig.~\ref{fig:fig3}\textbf{b},\textbf{c} the fast-Fourier-transformed (FFT) patterns of two selected regions, A and B, with the size of about $20\times 20\, \mathrm{nm}^2$. Focusing on domain A, we clearly observe superlattice reflections $\pm\bm{q}_1$ around forbidden fundamental reflections ($210$ and $2\mbox{-}10$). The coordinate system of this orthorhombic domain is depicted in the inset of Fig.~\ref{fig:fig3}\textbf{b}. The denoised inverse FFT amplitude obtained from the superlattice reflections clearly show an atomic modulation propagating along the $\bm{a}^*$ direction (Fig.~\ref{fig:fig3}\textbf{d}).

By contrast, the FFT pattern of domain B (Fig.~\ref{fig:fig3}\textbf{c}) shows superlattice reflections only at positions rotated by $240^\circ$ with respect to those in area A. Consistently, the corresponding real-space amplitude map (Fig.~\ref{fig:fig3}\textbf{e}) reveals a superlattice modulation propagating along this $240^\circ$-rotated direction relative to Fig.~\ref{fig:fig3}\textbf{d}. Unlike the XRD data in Fig.~\ref{fig:fig2}\textbf{f}, which averages over all domains because of the much larger sample volume probed, our TEM data resolves the nanoscale structure of the harmonic modulation and unambiguously establishes a single dominant propagation direction in each domain (single-$\bm{q}$ character).

\bigskip
\textbf{Sublattice-selective BDW orders satisfying the bond zero-sum constraint}\\
Guided by TEM observations, we refine the superstructure of \CRS{} using the XRD data at $300\,$K (BDW-I) in a $9 \times 1 \times 1$ supercell without imposing any constraints on the atomic positions beyond those required by the space group symmetry ($Pbmm$). Considering more than $100{,}000$ reflections, we achieve a high-quality refinement with $R(I>3\sigma) = 2.55\ \%$ and reasonable atomic displacement parameters (see Supplementary Section). The resulting superstructure corresponds to a transverse optical mode that primarily modulates interlayer bond lengths; we therefore term it the BDW order.

\begin{figure*}[t]
  \begin{center}
		\includegraphics[clip, trim=0cm 0cm 0cm 0cm, width=1.\linewidth]{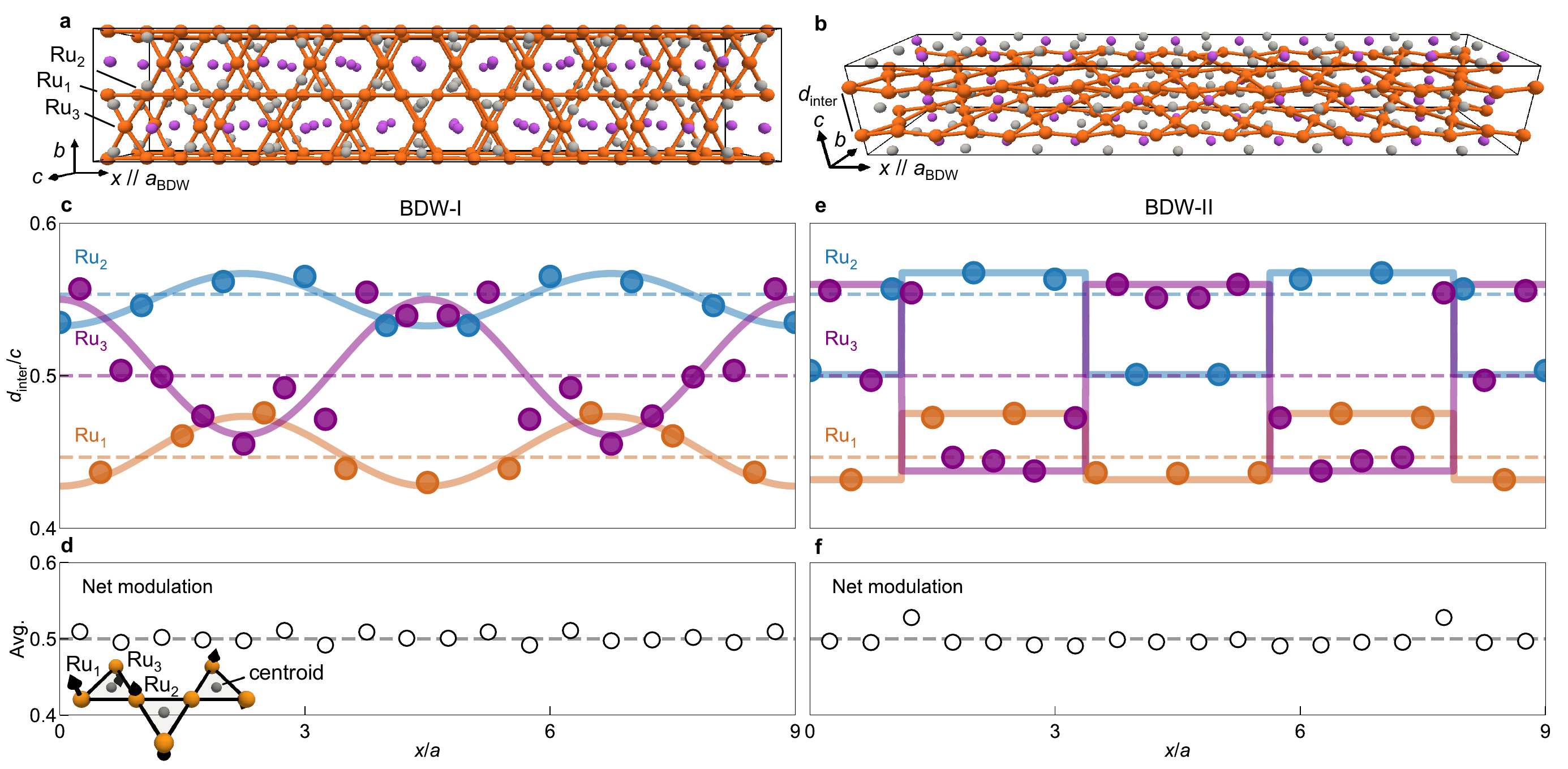}
    \caption[]{\textbf{Sublattice-selective BDW orders with zero net modulation per triangle.} \textbf{a},\textbf{b}, Top and side views of the refined $9\times 1\times 1$ superstructure at $300\,$K (BDW-I). \textbf{c},\textbf{e}, Interlayer bond length $d_\mathrm{inter}$ of Ru atoms, divided by the lattice constant $c$, as a function of the position $x$ along the $a_\mathrm{BDW}$ axis ($a_\mathrm{BDW}=9a$) in the $9 \times 1 \times 1$ supercell, obtained from superstructure refinement of BDW-I and BDW-II with the synchrotron XRD data. Orange, cyan, and purple denote Ru$_1$, Ru$_2$, and Ru$_3$ atoms, respectively. Dashed lines indicate $d_\mathrm{inter}/c$ in the parent stripe phase, with the purple dashed line marking the unbonded reference position ($d_\mathrm{inter}/c = 1/2$). Solid lines represent fitted sinusoidal modulations in panel~\textbf{c} and their square-wave approximations in panel~\textbf{e} with the primary modulation vector $\bm{q}_1 = (2/9, 0, 0)$. The Ru$_3$ sublattice forms the out-of-phase wave whose amplitude is twice larger than that of the other two waves. \textbf{d},\textbf{f}, Triangle-averaged $d_\mathrm{inter}/c$ (net modulation per triangle) assigned at the triangle centroids (gray spheres in the inset). The distance from the dashed line quantifies the degree of net bond neutrality on each triangle. The inset illustrates the sublattice-selective bond modulation relative to the reference $d_\mathrm{inter}$. Error bars are estimated to be several orders of magnitude smaller than the modulation amplitude and are therefore omitted (Methods).
    }
    \label{fig:fig4}
  \end{center}
\end{figure*}

Now we examine the details of BDW-I. Let $d_\mathrm{inter}$ describe the Ru-Ru bonds length between kagome layers. Figure~\ref{fig:fig4}\textbf{c} illustrates $d_\mathrm{inter}/c$ for each Ru sublattice in BDW-I as a function of the position $x$ along the $a_\text{BDW}$ axis ($a_\text{BDW}$ = 9$a$). Focusing on a single kagome layer of the parent phase (stripe, Fig.~\ref{fig:fig2}\textbf{b}), Ru$_1$ and Ru$_2$ make bonds between opposite kagome planes ($d_\mathrm{inter} = c/2 \pm \delta$), while Ru$_3$ remains unbonded as noted before ($d_\mathrm{inter} = c/2$). In BDW-I, Ru$_1$ and Ru$_2$ undergo a sinusoidal bond modulation, oscillating about their original bond lengths with amplitude $\delta$. By contrast, Ru$_3$ develops a bonding wave centered at $d_\mathrm{inter}=c/2$ with amplitude $2\delta$, out of phase with the modulations of Ru$_1$ and Ru$_2$. Importantly, the BDW patterns depend on the kagome sublattices --- hence the term sublattice-selective --- and satisfy the bond zero-sum constraint on each triangle, $\delta + \delta - 2\delta = 0$, analogous to the zero-sum constraint in frustrated antiferromagnets (Eq.~\ref{eq:1}). The bond zero-sum constraint is reflected in the negligible deviation of the net modulation per triangle from the pristine bond length $c/2$ (Fig.~\ref{fig:fig4}\textbf{d}).

Similarly, we refine the superstructure at $50\,$K (BDW-II) in the same space group and the supercell size as BDW-I with high quality: $R(I>3\sigma) = 4.18\ \%$. In contrast to BDW-I, BDW-II exhibits discrete levels of $d_\mathrm{inter}$ (Fig.~\ref{fig:fig4}\textbf{e}). Each sublattice forms an anharmonic square BDW, in line with the observation of higher-order harmonics of $\bm{q}_1$ in Fig.~\ref{fig:fig2}\textbf{g}. Once again, Ru$_3$ modulates out of phase to preserve the local bond zero-sum constraint (Fig.~\ref{fig:fig4}\textbf{f}). We reiterate that no positional constraints were imposed during the structure refinements; nevertheless, the obtained superstructures spontaneously obey the bond zero-sum constraint. 

\begin{figure*}[t]
  \begin{center}
		\includegraphics[clip, trim=0cm 0cm 0cm 0cm, width=0.8\linewidth]{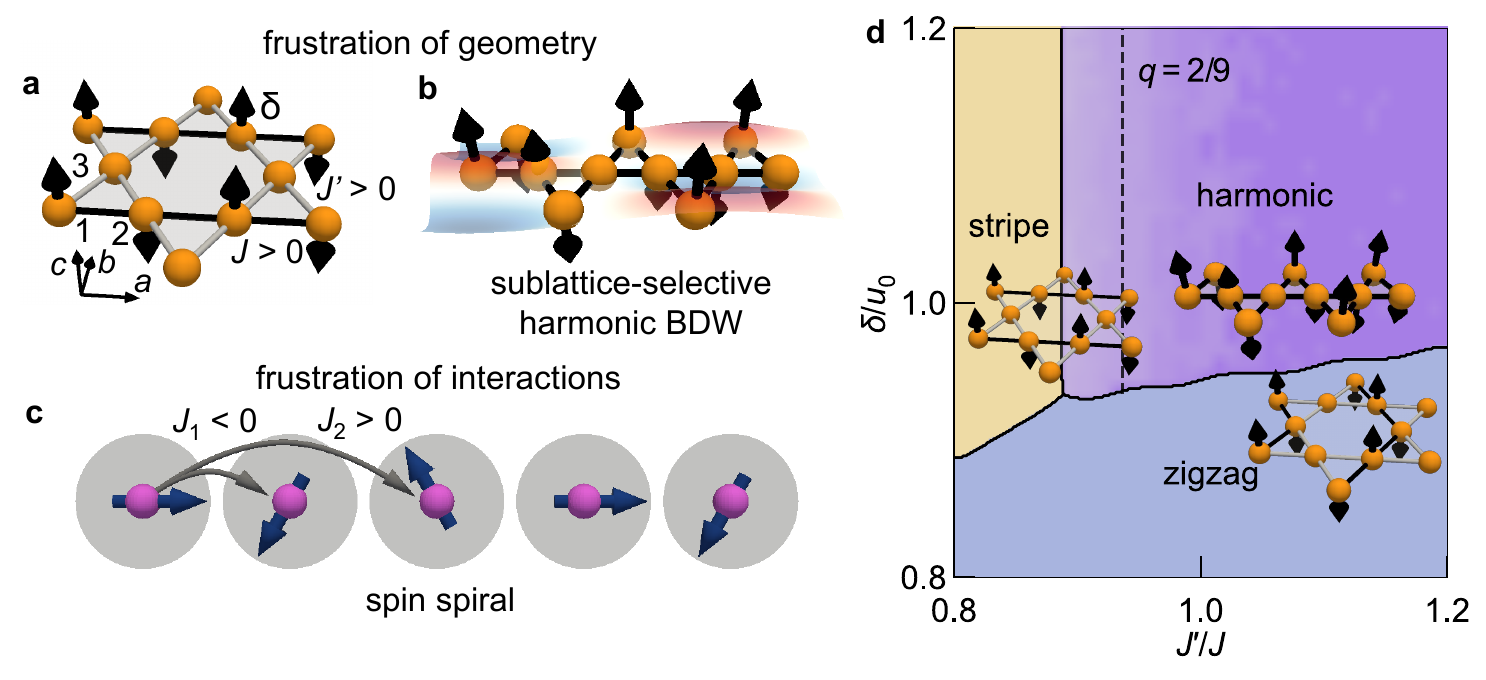}
    \caption[]{\textbf{Anisotropic nearest-neighbor soft-Ising model on the kagome lattice.} \textbf{a}, Schematic of the anisotropic pseudospin model derived from an expansion of the Coulomb repulsion. $J$ ($J'$) denote the nearest-neighbor pseudo-exchange interactions between sublattices $1$–$2$ ($1$–$3$ and $2$–$3$). The parent stripe phase with bond modulations of $\pm\delta, 0$ is illustrated, and the inequivalence of $J$ and $J'$ arises from the difference in bond lengths. The interlayer bond modulation $\bm{u}_i(\bm{r})$ on sublattice $i$, defined relative to the stripe phase, is regarded as a pseudospin variable. \textbf{b}, Sublattice-selective harmonic BDW order induced by anisotropic exchange interactions and geometric frustration. The three sublattices on the kagome lattice exhibit the sinusoidal bond modulations $\bm{u}_1(\bm{r})=\bm{u}_2(\bm{r}) =-1/2\bm{u}_3(\bm{r}) = u_0\bm{e}_z\cos(qx)$ to satisfy the local bond zero-sum constraint.
    \textbf{c}, Spin spiral arising from competing nearest-neighbor ferromagnetic ($J_1<0$) and next-nearest-neighbor antiferromagnetic ($J_2>0$) interactions. Geometric frustration enforces a sublattice-selective texture, leading to effective frustration of exchange interactions: $E_{\text{ex}} = u_0^2\left(- 4J'\cos{q/4} + J\cos{q/2}\right)$.  \textbf{d}, Phase diagram of the anisotropic pseudospin model. The color gradient in the harmonic BDW phase indicates its wavenumber $q$, with the thickest shading corresponding to $q = 0$ and the thinnest to $q = 1/2$. The parameter regime that reproduces the observed $q = 2/9$ harmonic BDW order is highlighted by a broken line. The ground state is determined within the variational manifold of bond-ordered states shown in Fig.~\ref{fig:fig1}\textbf{f}–\textbf{i}, all of which are observed experimentally. Inclusion of higher-order interactions stabilizes the square BDW order (see Methods).
    }
    \label{fig:fig5}
  \end{center}
\end{figure*}

\bigskip
\textbf{Anisotropic soft-Ising model on the kagome lattice.}\\
To understand the origin of the sublattice-selective BDW orders, we derive a pseudospin model from the Coulomb repulsion between $d_{z^2}$ orbitals. We start from the parent stripe phase, where the in-plane distances of Ru$_1$-Ru$_3$ and Ru$_2$-Ru$_3$ are equal but shorter than that of Ru$_1$–Ru$_2$, resulting in anisotropic Coulomb interactions (Fig.~\ref{fig:fig5}\textbf{a}). Defining $\bm{u}_i$ as the additional bond modulation on sublattice $i$, we expand the Coulomb repulsion and obtain the total next-nearest neighbor interaction per triangle, 
\begin{equation}
E = E_0 - (\bm{h}+\bm{h}')\cdot(\bm{u}_1 - \bm{u}_2) - (K + K')(\bm{u}_1^2 + \bm{u}_2^2) - 2K'\bm{u}_3^2 + J\bm{u}_1\cdot\bm{u}_2 + J'(\bm{u}_2\cdot\bm{u}_3 + \bm{u}_3\cdot\bm{u}_1),
\end{equation}
where $E_0$ is the energy of the parent phase (see Supplementary Section for details). Interpreting $\bm{u}_i$ as a pseudospin, the linear term acts as an effective field, the quadratic terms as single-ion anisotropy, and the bilinear terms as exchange interactions, whose coefficients may be renormalized by the elastic cost of bond modulations.
 
The sublattice-selective harmonic BDW order satisfies $\bm{u}_1(\bm{r}) = \bm{u}_2(\bm{r}) = -1/2 \bm{u}_3(\bm{r}) = u_0\cos (qx_i) \bm{e}_z$ (Fig.~\ref{fig:fig5}\textbf{b}). The resulting exchange interaction becomes translationally invariant, as bond-position-dependent terms average to zero over the lattice. The exchange interaction is then $E_{\text{ex}} = u_0^2\left(- 4J'\cos{q/4} + J\cos{q/2}\right)$ implying effectively ferromagnetic coupling between sublattices 1(2) and 3 and antiferromagnetic coupling between sublattices 1 and 2. Although all interactions between pseudospins are antiferromagnetic in origin, geometric frustration and the resulting zero-sum constraint enforce a competition between effective interactions. The expression for $E_\mathrm{ex}$ is well-known to create incommensurate spiral chains (Fig.~\ref{fig:fig5}\textbf{c}). Minimization yields the observed $q = 2\pi\times 2/9$ of BDW-I for $J'/J = \cos(\pi/9) \approx 0.94$.

To assess the robustness of the harmonic BDW state in the model, we construct the ground-state phase diagram using a variational approach within the manifold shown in Fig.~\ref{fig:fig1}\textbf{f}-\textbf{i}. The harmonic BDW order is stable over a parameter range consistent with the analytical result and competes with other bond-ordered states (Fig.~\ref{fig:fig5}\textbf{d}). We also show that further-neighbor interactions stabilize the square BDW order, in line with their enhancement as electronic screening weakens at low temperatures (see Methods).

\bigskip
\textbf{Discussion}\\ 
Long-period superlattices typically manifest without genuine sublattice selectivity. Even in well-studied spin systems, modulations on different sublattices are typically identical or symmetry-equivalent (Fig.~\ref{fig:fig1}\textbf{e}). By contrast, the BDW orders discovered here combine long-period structural modulation with explicit sublattice selectivity, even though the low orthorhombic symmetry lifts the degeneracy of the kagome sublattices. This shows that the sublattice-selective modulation is not a trivial consequence of symmetry but rather arises from a local bond zero-sum constraint that minimizes inter-bond Coulomb repulsion on the kagome lattice, akin to neutrality constraints in frustrated spin and charge systems. While the square BDW order observed here resembles commensurate phases in devil’s staircases~\cite{von-Boehm1979-ju,Fisher1980-jd,Bak1982-hj,Selke1988-df}, the two-dimensional nature of the kagome lattice, together with the sublattice-selective bond modulation, places the BDW orders beyond the standard framework of quasi-one-dimensional systems with anharmonic atomic or spin modulations~\cite{Aramburu2006-nu,Kruger2009-pq,Boucher1996-hm,Leist2008-rm}. More broadly, the persistence of these BDW orders above room temperature establishes a new route for implementing frustration physics in quantum materials, extending the scope of geometric frustration beyond spin and charge degrees of freedom that commonly manifest only in low temperatures. Looking forward, we anticipate generalizing the bond zero-sum constraint physics to other frustrated lattices and harnessing the interplay between bond order and the electronic structure, lattice dynamics, superconductivity, and topology of kagome metals~\cite{Yin2022-ps,Checkelsky2024-ik,Di-Sante2026-lx}.

\newpage

\let\emph\relax
\bibliography{ref}

\newpage

\newpage

\begin{center}
\Large{Methods}
\end{center}

\textbf{Crystal growth}\\
Polycrystalline and single-crystalline samples of \CRS{} were synthesized by the arc melting technique in a high-purity argon atmosphere.
To avoid the competing impurity CeRu$_2$Si$_2$, we added excess Ru to the starting materials~\cite{Barz1980-ep}. Small single crystals, approximately $50\,\mu\mathrm{m}$ in size and suitable for SXRD measurements, were obtained by mechanical fragmentation of the ingots.

\bigskip
\textbf{Synchrotron X-ray diffraction}\\
Synchrotron XRD measurements were carried out at beamline BL$02$B$1$ of the SPring-$8$ synchrotron radiation facility (Japan). Diffraction patterns were collected using a CdTe PILATUS area detector. The sample temperature was controlled by nitrogen gas flow for temperatures above $100\,$K and by helium gas flow for temperatures below $100\,$K. Integrated intensities were processed using \textsc{CrysAlisPro}~\cite{Agilent-Technologies-Ltd2014-hl}, with equivalent reflections averaged. Structural refinements were performed using \textsc{Jana2006}~\cite{Petricek2014-pc}. As for structural refinements for BDW-I, we use a dataset from an independent measurement at the same temperature and in the same single crystal (Supplementary Section). For the temperature dependence of XRD intensities in Fig.~\ref{fig:fig2}\textbf{d}, the integrated intensity of each reflection was obtained by subtracting a background from the total counts within the peak region. Statistical uncertainties were then estimated assuming Poisson counting statistics. For space-group descriptions (for example, \(Ibmm\)), the crystallographic axes were retained as in the pristine hexagonal structure and were not transformed to the conventional setting (for example, \(Imma\)) to enable direct comparison. The data at $500\,$K in Fig.~\ref{fig:fig2}\textbf{c} were collected from the same single crystal in an independent measurement, with a different UB matrix and overall intensity scale than those in Fig.~\ref{fig:fig2}\textbf{f},\textbf{g}. The uncertainty of the refined atomic coordinate \(Z\) was estimated as $\Delta Z = Z\sqrt{(\Delta c/c)^2 + (\Delta z / z)^2}$, where $\Delta c$ and $\Delta z$ denote the uncertainties of the lattice constant obtained from unit-cell determination and the fractional atomic coordinate from structural refinement, respectively. In our case, typical values of $\Delta c / c$ and $\Delta z / z$ are on the order of $10^{-5}$ and $10^{-4}$ (Supplementary Section), respectively, which are much smaller than the modulation amplitudes shown in Fig.~\ref{fig:fig4}\textbf{c}-\textbf{f}. Error bars are therefore omitted.

\bigskip
\textbf{Transmission electron microscopy}\\
High-resolution transmission electron microscopy (HRTEM) was carried out using a transmission electron microscope (JEOL JEM-$2800$) operated at $200$ kV, equipped with a OneView camera (Gatan Inc.). TEM specimens were prepared by crushing the samples, followed by sonication in isopropyl alcohol, and depositing a few drops of the suspension onto TEM grids. All measurements were performed at room temperature. Because the temperature is close to $T_\text{BDW-II} \sim 300\,$K, $\bm{q}_2$ and $\bm{q}_1+\bm{q}_2$ superlattice reflections are weak and invisible.

TEM images were analyzed using fast Fourier transforms (FFT) to examine superlattice features in reciprocal space. Fourier amplitudes are displayed on a logarithmic scale. For the inverse FFT, superlattice reflections were selected in Fourier space according to their symmetry-allowed positions, while fundamental Bragg reflections and background contributions were excluded. A smooth Gaussian mask was applied around the selected superlattice reflections to minimize truncation artifacts, and the inverse FFT was performed to reconstruct the associated real-space modulation. The amplitude of the reconstructed image was lightly Gaussian-smoothed for visualization.

\bigskip
\textbf{Pseudospin model calculations}\\
Model calculations were performed on a two-dimensional chain extending along the $a$ direction and consisting of $10{,}080$ triangles and inverted triangles (Fig.~\ref{fig:fig5}\textbf{a}). This reduction is justified because all ansatz states considered have identical configurations along the $b$ axis. For each point in the parameter space defined by $\delta/u_0 \times J'/J$ on a $40\times40$ grid, the modulation wave number $q$ of the BDW orders was optimized. The phase diagram was then constructed by comparing the total energies within the variational manifold.

At elevated temperatures, thermal carriers screen the Coulomb interactions, rendering long-range interactions less effective. Therefore, only the anisotropic nearest-neighbor interactions are considered for BDW-I. On the other hand, for BDW-II, third-nearest neighbor ($3$NN) interactions ($J_3$) were included as the minimal additional ingredient required to stabilize the anharmonic square BDW order (Supplementary Section). In phase diagrams with $J_3\neq 0$, $q=1/10$ and $q=1/3$ square BDW orders appear. By contrast, $J_3$ destabilizes the harmonic BDW orders because the effective ferromagnetic coupling between sublattices 1(2) and 3, required for the harmonic BDW orders (Fig.~\ref{fig:fig5}\textbf{a}-\textbf{c}), is absent for $3$NN channels. Rather, antiferromagnetic interactions become dominant and prefer $q=1/2$.
\bigskip

\textbf{First-principles calculations}\\
Density functional theory calculations were performed using the Vienna Ab initio Simulation Package (VASP) with the projector augmented wave (PAW) method~\cite{Kresse1994-ct,Kresse1996-ij,Kresse1996-uh}. The generalized gradient approximation (GGA) with Perdew-Burke-Ernzerhof (PBE) was adopted for the exchange-correlation energy functional~\cite{Perdew1996-ih}. A plane-wave energy cutoff of $500$~eV was used. During structural optimization, the energy convergence criteria were set to $10^{-8}$~eV and the reciprocal space was sampled using the Monkhorst-Pack scheme with a $k$-mesh of $9\times 9\times 11$. The structure was fully relaxed before the phonon calculation with a force convergence criteria of $10^{-3}$~eV/A. The phonon spectra and force constants by VASP are extracted via PHONOPY code with density functional perturbation theory (DFPT)~\cite{Togo2015-tw}, using a $2\times 2\times 2$ supercell with a $\Gamma$-centered $4\times 4\times 6$ $k$-mesh.

\bigskip
\textbf{Acknowledgements}\\
We thank M. Jovanovic, Y. Tanaka, K. Kakurai, K. Shimizu, A. Kajihara, and C. Pollak for fruitful discussions. X.Y. acknowledges Dr.~Yi-Ling Chiew for her technical support. This work was supported by JSPS KAKENHI Grants No. JP22K20348, No. JP23H05431, No. JP23K13057, No. JP24H01607, No. JP24H01604, No. JP25K17336, JP24K17006, and 24H01644, as well as JST CREST Grant No. JPMJCR20T1 (Japan) and JST PRESTO Grant No. JPMJPR259A and JST FOREST Grant No. JPMJFR2238, and No. JPMJFR2362 (Japan). It was also supported by JST as part of Adopting Sustainable Partnerships for Innovative Research Ecosystem (ASPIRE), Grant Number JPMJAP2426. Work at Princeton was supported by the David and Lucille Packard foundation, the Princeton Catalysis Initiative (PCI), and by the Air Force office of Scientific Research under award number FA9550-24-1–011. P.R.B. acknowledges SNSF Postdoc.~Mobility grant P500PT$\underline{}$217697 for financial assistance.
X. Y. is supported by the RIKEN TRIP program. The synchrotron single-crystal X-ray experiments were performed at BL02B1 of SPring-8 with the approval of RIKEN (Proposal No. 2024A1760, 2024B1839, and 2024B2010). 

\bigskip
\textbf{Data and code availability}\\
The data are available from the corresponding authors upon reasonable request. 

\bigskip
\textbf{Author contributions}\\
M.H. and L.M.S. conceived the project. R.M. grew and characterized the single crystals with support from R.Y. R.Y and R.N. made polycrystalline samples with support from L.M.S. R.M., S.K., R.Y., P.R.B, M.H., and Y.N. performed single crystal XRD measurements at beamline BL02B1 of SPring-8. R.M. analyzed the XRD data with guidance from S.K. and T.-h.A. X.Y. conducted transmission electron microscopy (TEM) measurements, and X.Y. and R.M. analyzed the TEM data. R.M. constructed the pseudospin model and performed analytical calculations. R.M. performed and analyzed numerical calculations of the pseudospin model with guidance from Y.M. X.F. performed and analyzed first-principles calculations. R.M. wrote the manuscript in collaboration with M.H. and S.K.; all authors discussed the results and commented on the manuscript.

\bigskip
\textbf{Competing interests}\\
The authors declare no competing interests.

\clearpage
\begin{center}
\Large{Main Text Figures}
\end{center}

\clearpage

\clearpage

\clearpage

\clearpage

\end{document}